\documentclass[sigplan,10pt,table]{acmart}
\renewcommand\footnotetextcopyrightpermission[1]{} %

\usepackage[utf8]{inputenc}
\usepackage[english]{babel}

\usepackage{multirow, multicol, booktabs, tabulary, tabu, longtable, array, varwidth}
\usepackage{placeins, lipsum}
\usepackage[flushleft]{threeparttable}
\usepackage{enumerate}
\usepackage[labelfont=bf]{caption}
\usepackage{pifont} %
\newcommand{\cmark}{\ding{51}}%
\usepackage[font=small,labelfont=bf]{caption}
\newcommand{\xmark}{\ding{55}}%

\usepackage{enumerate}

\usepackage[inline]{enumitem}

\usepackage{amsmath, amsfonts, amssymb, amsthm, nicefrac}
\usepackage{siunitx}
\theoremstyle{definition}

\usepackage{listings}

\usepackage[scaled=0.82]{beramono}

\usepackage[]{cleveref} %
\crefname{appsec}{Appendix}{Appendices}
\crefformat{section}{\S#2#1#3}
\crefformat{subsection}{\S#2#1#3}
\crefformat{subsubsection}{\S#2#1#3}
\crefformat{equation}{(#2#1#3)}
\crefrangeformat{equation}{(#3#1#4--#5#2#6)}
\crefmultiformat{equation}{(#2#1#3)}{ and~(#2#1#3)}{, (#2#1#3)}{ and~(#2#1#3)}
\crefformat{figure}{Fig.~#2#1#3}
\crefrangeformat{figure}{Figs. #3#1#4--#5#2#6}
\crefmultiformat{figure}{Figs.~#2#1#3}{ and~#2#1#3}{, #2#1#3}{ and~#2#1#3}
\crefname{algocf}{Alg.}{Algs.}
\Crefname{algocf}{Algorithm}{Algorithms}
\crefformat{table}{Table~#2#1#3}
\crefrangeformat{table}{Tables~#3#1#4--#5#2#6}
\crefmultiformat{table}{Tables~#2#1#3}{ and~#2#1#3}{, #2#1#3}{ and~#2#1#3}

\usepackage{textcomp}
\usepackage[shortcuts,acronym]{glossaries}
\usepackage{soul, footnote, xargs}
\usepackage[colorinlistoftodos,prependcaption,textsize=tiny]{todonotes}
\usepackage{balance}

\usepackage{multirow, multicol, booktabs, tabulary, tabu, longtable, array, varwidth}
\usepackage{placeins, lipsum}
\usepackage[flushleft]{threeparttable}

\usepackage{graphicx}
\usepackage{subcaption}
\usepackage{tikz, epstopdf, stfloats, bbding, capt-of}
\usepackage{algorithmic}
\usepackage[ruled, linesnumbered]{algorithm2e}

\SetCommentSty{mycommfont}

\usepackage{microtype}
\usepackage[T1]{fontenc}

\definecolor{gray}{rgb}{0.5,0.5,0.5}
\definecolor{mauve}{rgb}{0.58,0,0.82}
\definecolor{mygreen}{rgb}{0,0.6,0}
\newcommand{\overload}{overload\xspace}
\newcommand{\sysname}{Kaleidoscope\xspace}
\newcommand{\lustre}{{Lustre}\xspace}

\newcommand{\ost}{{OSD}\xspace}

\newcommand{\mgs}{{MGS}\xspace}

\newcommand{\lnet}{{LNET}\xspace}

\newcommand{\storage}{PetaStore\xspace}

\begin{document}

\date{}

\title{Live Forensics for Distributed Storage Systems}

\author{Saurabh Jha$^1$, Shengkun Cui$^1$, Tianyin Xu$^1$, Jeremy Enos$^{2}$, Mike Showerman$^{2}$,  Mark Dalton$^3$, Zbigniew T. Kalbarczyk$^1$, William T. Kramer$^{1,2}$,  Ravishankar K. Iyer$^1$ }
\affiliation{%
 \institution{$^1$University of Illinois at Urbana-Champaign, $^2$National Center for Supercomputing Applications,$^3$ Cray Inc}
}

\maketitle
\thispagestyle{empty}
\subsection*{Abstract}
We present \sysname{} an innovative system that supports live forensics for 
    application performance problems caused by either 
    individual component failures 
    or resource contention issues in large-scale distributed 
    storage systems.
The design of \sysname{} is driven by our study of I/O failures observed 
    in a peta-scale storage system anonymized as \storage{}. 
\sysname{} is built on three key features: 
1) using temporal and spatial differential observability for end-to-end performance monitoring of I/O requests, 
2) modeling the health of storage components as a stochastic process 
using domain-guided functions that accounts for path redundancy and uncertainty in measurements, 
and, 3) observing differences in reliability and performance metrics between similar types of healthy and unhealthy 
components to attribute the most likely root causes.
We deployed \sysname{} on \storage{} and our evaluation shows that 
\sysname{} can run live forensics at 5-minute intervals 
    and pinpoint the root causes of 95.8\% of real-world performance issues,
with negligible monitoring overhead.

\section{Introduction}

Large-scale storage services are typically implemented on top of clusters of servers and disk arrays to provide high performance (e.g., load balancers and congestion control) as well as high availability (e.g., RAID, and active-active high availability server pairs). 
Component failures~\cite{khan2011search,wan2010s2,kbp:12:rec, Ford:2010, gunawi2018fail} and resource contention~\cite{cao2017performance,kim2017enlightening} are chronic problems that lead to I/O timeouts and slowdown in such systems.  State-of-the-art solutions focus on reliability failures (e.g., DeepView~\cite{Zhang:2018} and Panorama~\cite{Huang:2018}) and hence, do not attempt to distinguish between resource contention and component failures in storage systems. We assert that knowing whether a problem is due to resource contention or component/node/subsystem failure is critical in effectively coordinating a recovery strategy. For example, in the case of component failures, an immediate repair action must be taken to avoid failures during fail-over and recovery~\cite{ji2019automatic,brown2001embracing,ma2015raidshield}.  
In the case of resource contention, a solution may involve load balancing and throttling of excessive I/O requests 
    generated by applications (as well as restructure the code).

A combination of component failures and contention issues 
    significantly degrades application performance in production settings (see \cref{sec:understanding}). This paper uses  a combination of proactive monitoring  and machine learning to jointly address the above issues. We have incorporated the proposed techniques into an automated tool called \sysname{}. Our tool has been demonstrated in live traffic on a production system to 1) locate components (e.g., data servers and RAID devices) causing I/O bottlenecks (i.e., I/O slowdown or timeouts), 2) differentiate between a reliability failure and a resource contention issue, and 3)  quantify the negligible impact on the system performance while delivering high precision and recall, as discussed later in this section.

To support  failure detection and live forensics, \sysname uses the following novel techniques:
\begin{itemize}[noitemsep,nolistsep,leftmargin=*] 
\item {\bf Proactive monitoring.} \sysname{} monitors the end-to-end performance of a storage system using Store-Pings, a set of monitor primitives that cover all the storage operations involved in serving a client's I/O requests (e.g., create, read, write, and delete files). Store-Ping monitors are strategically placed to provide both spatial and temporal differential observability in real time.

\item {\bf Modeling and inferring component health.} The health of a component in a storage system (e.g., a metadata server or a RAID device) is modeled as a stochastic process that accounts for uncertainty (due to performance variability and asynchrony) as well as non-determinism in distributed storage systems. We built a system model by using the factor graph (FG) formalization, which infers component health by ingesting the monitoring data collected by Store-Pings. The inference on the model allows \sysname{} to localize unhealthy components in near real-time.
 
\item {\bf Methods to determine the cause of I/O failures.} A set of statistical methods (including a local outlier factor~\cite{Breunig:2000:LID:335191.335388} algorithm run using data on server load, disk load, and disk bandwidth utilization) and clustering~\cite{taerat2011baler} of storage system error logs are used to distinguish between component failures and resource overloads. The statistical methods are based on comparison of reliability and performance metrics (such as the number of active processes on a data server) as they are collected for healthy and unhealthy components. Note that the distinction between healthy and unhealthy components is provided by the FG-based model discussed above.

\end{itemize}

\paragraph{\bf Deployment.} \sysname{} has been deployed on PetaStore, a 36 PB production system, 
which employs the Lustre file system~\cite{lustre}. 
Lustre is used by more than 70\% of the top 100 supercomputers~\cite{lustretop100} and
is offered by cloud service vendors such as Amazon and Azure~\cite{amazonFsx,AzurePVFS}.
Its design resembles that of many other 
object-based POSIX storage systems, such as IBM GPFS~\cite{schmuck2002gpfs}, BeeGFS~\cite{heichler2014introduction}, Ceph~\cite{weil2006ceph}, and  GlusterFS~\cite{osti_1048672}.

\paragraph{\bf Monitoring overhead.} Store-Ping-based monitors have been deployed on \storage{} for two years. 
The monitors measure the completion times of 5,382 I/O requests per minute and cover every I/O path from any client to a RAID device. Since Store-Pings actively collect data at 60-second intervals, we measured the overhead introduced by Store-Ping monitors on the production system and found the overhead to be less than 0.01\% on the peak I/O throughput of \storage{}.

\paragraph{\bf Forensic effectiveness.}
We used two years of monitoring data collected by Store-Ping monitors to evaluate the effectiveness of 
    \sysname{}'s live forensics.
The evaluation is based on
    843 production issues identified and resolved by the \storage{} operators in this period as the ground truth.
Overall \sysname :
\begin{itemize}[noitemsep,nolistsep,leftmargin=*] 
\item correctly localizes the component failures (e.g., a specific data server or a RAID device)  and resource overloads for 99.3\% of cases.
\item accurately identifies likely root cause for 95.8\% of cases, i.e. disambiguates between resource contention and component failures.
\item is configured, in \storage{}, to run data collection at one-minute intervals and produce forensics at 5 minute intervals. 
Our results indicate that \sysname can collect and produce forensics using 100 monitors at 30 second interval with marginal impact of 2.42\% on the peak throughput.
\end{itemize}

\section{Motivation and Goals}
\label{sec:motivation}

\begin{figure}[t]
    \centering
    \includegraphics[width=0.4\textwidth]{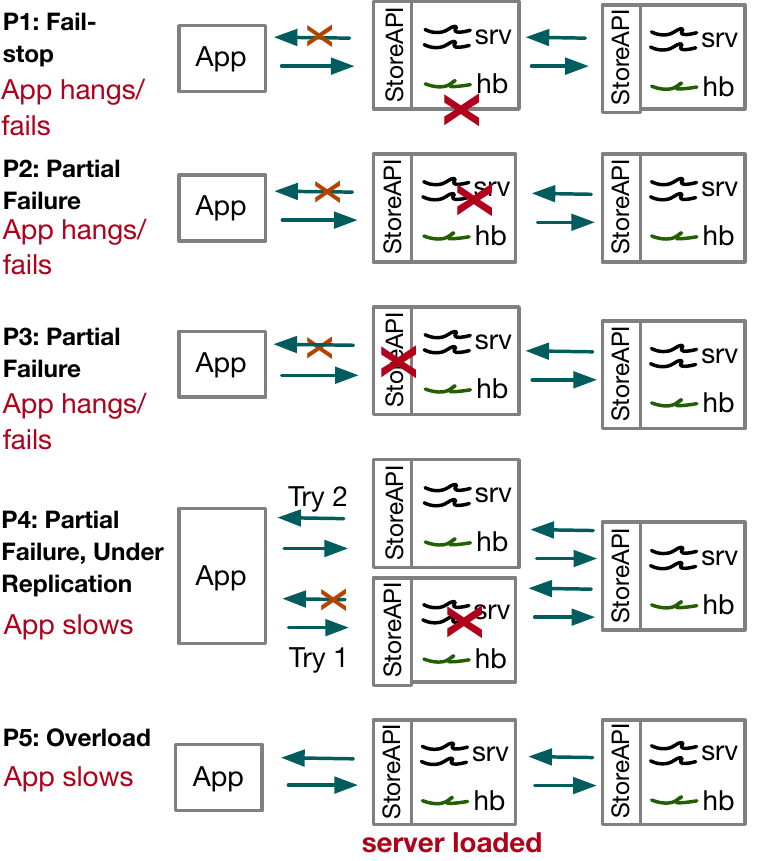}
    \vspace{-0.4cm}
    \caption{\small Common patterns of I/O failures. Notation: "hb" is  heartbeat process, "srv" is service process; and each box represents the storage components (e.g., data servers).}
    \label{fig:io_failures}
    \vspace{-0.3cm}
\end{figure}

We address the following failure patterns~\footnote{\cite{avizienis2004basic} provides a taxonomy of failure patterns.} that are frequently experienced 
    in large-scale distributed storage systems~\cite{Huang:2017, bairavasundaram2008analysis,Ford:2010} such as \storage{} (\S\ref{sec:understanding}),
    as illustrated in Figure~\ref{fig:io_failures}:
\begin{itemize}[noitemsep,nolistsep,leftmargin=*] 
    \item {\bf P1}: Fail-stop failures of an entire storage component (e.g., crash of data server)~\cite{vanRenesse:2009:GFD:1659232.1659238,Chandra:1996} ;
    \item {\bf P2}: Fail-stop failures of a service process or a thread in the storage component~\cite{leners2011detecting};
    \item {\bf P3}: Gray failures~\cite{Huang:2018,Huang:2017} that are visible to the application (e.g., an I/O timeout) but not to the failure detectors. They occur because of differential observability;
    \item {\bf P4}: Fail-slow failures in which the built-in redundancy masks the failures 
        from applications, 
        but results in increased I/O response time~\cite{muntz1990performance,wu2009workout,Wu:2007:PUF:1362622.1362687,1410735};
    \item {\bf P5}: Fail-slow failures due to high loads or contention in which the storage system takes more time to service an I/O request because of contention but does not lead to I/O timeout~\cite{6337777,Kraft2013}. %
\end{itemize}

Our goal is to detect those issues and provide live forensic support 
    to localize unhealthy components and pinpoint root causes.
In large-scale storage systems (\S\ref{sec:study}), failures with the above different patterns 
    often manifest through similar symptoms known as {\it I/O failures}; an I/O request
    is considered to have "failed" if it has not been serviced within an expected amount of time.
The difficulty of distinguishing causes from symptoms often results in a lengthy
    troubleshooting process and significantly prolongs failure recovery. (different failure patterns
    require different recovery strategies.)    

\begin{table}%
\begin{tabular}{|l|l|l|l|l|l|l|}
\hline
\multirow{2}{*}{\bf Detection method}  & \multicolumn{5}{c|}{\bf Failure Patterns} \\
                            & P1  & P2  & P3  & P4 & P5 \\ 
\hline
\hline
Basic heartbeat             &  \cmark    &   \xmark   &   \xmark   &   \xmark  &  \xmark    \\ \hline
Service-aware heartbeat     &  \cmark   & \cmark  & \xmark     & \cmark    & \xmark     \\ \hline
Client-view observation     &  D &     D &     D & \xmark    & \xmark    \\ \hline
\sysname                    &  \cmark   & \cmark  & \cmark     & \cmark    & \cmark     \\ \hline
\end{tabular}
\caption{\small Failure patterns and the capability of different types of failure detection.
    \cmark\xspace: can detect the failure; 
    \xmark\xspace: cannot detect the failure;
    D: the failure is detected only if the client is affected.}~\label{tab:monitor_perf}
\end{table}

\subsection{The State of the Art}
\label{sec:comparison}

In the following, we show that the state-of-the-art approaches are fundamentally 
    limited in dealing with the aforementioned failure patterns that commonly
    occur in the field, especially in distributed storage systems. 
\cref{tab:monitor_perf} summarizes three different classes of existing approaches
    and compares them with our \sysname approach. 
We focus on failure detection in the rest of this section, but we would like 
    to note that \sysname{} is beyond a failure detector;
    it provides end-to-end live forensics to pinpoint root causes.
    
We classify existing failure detection methods into three categories: 
    (1) {\it basic heartbeats}~\cite{vanRenesse:2009:GFD:1659232.1659238,Hayashibara:2004} in
        which a dedicated heartbeat process running on each component indicates 
        the status (\texttt{UP} or \texttt{DOWN}) of the component;
    (2) {\it service-aware heartbeats}~\cite{1405971,Leners:2015:TUD:2741948.2741976, leners2011detecting,794001} in which the 
        heartbeat process validates the liveness and functional correctness
        of the services~\footnote{Any basic heartbeat can be converted to service-aware heartbeats.}; 
    and (3) {\it client view observations}~\cite{Zhang:2018, Huang:2018}
        in which failures are detected based on the observations of the clients.
        
As shown in Table~\ref{tab:monitor_perf}, heartbeat-based methods are insufficient. 
    Ideally, service-aware heartbeats can report the precise status of a component;
    however, it is hard to check the fine-grained 
    functionality of each component, since they are typically implementation-specific.
Moreover, a component could depend on many other components,
    making it hard to scale.
Client-view observation can close the gaps of differential observability, but 
    it cannot deal with failure patterns that have not manifested as external, client-perceivable issues such as Patterns 4 and 5.
On the other hand, our study in \S\ref{sec:study} shows that Patterns 4 and 5 
    are among the biggest threats to availability and performance
    in large-scale distributed storage systems.
\sysname{} is designed to address all the failure patterns in Table~\ref{tab:monitor_perf}.

We argue that the common practices of failure detection (as discussed above) 
    do not consider the inter-relation between reliability and performance.
However, as we demonstrate in this paper, the ability to differentiate between component failures 
    and resource overloads is of vital importance.

\subsection{Principles}
\label{sec:principle}

To meet the requirements of proactive failure detection 
    and live forensics, we employed several design principles that are 
crucial to the success of \sysname{}:

\begin{itemize}[noitemsep,nolistsep,leftmargin=*] 
\item {\bf Observability.}
We focus on enhancing observability to achieve fast failure detection and
	live forensics from both the client's and the server's views.
The client-view helps us detect user-perceived I/O failures,
	while the server-veiw helps us pinpoint the root causes 
	on the server side. 
The view points are 
	both spatial (across clients) and 
	temporal (in time series).

\item{\bf Dealing with uncertainty.} Failure detection and forensics
 	have to take into account the uncertainty introduced by the complexity of the
	production environment (e.g., random path selection) 
	and noise associated with measurements (e.g., transient delay in I/O packets).
	
\item{\bf Automation.}
We aim to create a fully automated and unsupervised system that can work with
    massive and imperfect production data.
Tools are less useful if they require manual classification and reasoning.

\item{\bf Localization support.}
    We aim to localize the unhealthy components that lead to the detected
	I/O failures. We find that aggregation of the client-side observations offers
	great opportunities for effective localization, because a misbehaving component often
	affects multiple clients.

\item{\bf Identifying root causes.}
We find a strong need to pinpoint the root cause of I/O failures in real time,
	because recovery strategies are based on root causes.
Distinguishing a) I/O failures caused by component failures
	and b) resource overloads would be particularly useful.

\item{\bf Low overhead.} Fault tolerance and forensics
	cannot affect or interfere with the performance of the normal workloads.
\end{itemize}

\section{Understanding Failures} 
\label{sec:understanding}

\sysname{} is driven by insights from the daily operations of a petascale 
    distributed storage system anonymized as \storage{}
    and the analysis of production failures at \storage{}.

\subsection{\storage{}}
\label{sec:petasore}

\storage{} is designed for large-scale, high-performance computing with I/O intensive 
    workloads, such as machine learning and large-scale simulations.
\storage{} consists of 6 management servers, 6 metadata servers, 420 data servers, and 582 I/O forwarding nodes (LNET nodes). The storeage servers in \storage{} are connected via an 
	internal Infiniband network, serving
    28,000+ computing nodes (as clients).
	
\storage{} uses	the \lustre distributed file system to manage 36 PB disk space across 17,280 disk devices. 
The disks are arranged in grid RAID~\cite{holland1992parity}, and referred as object storage devices (OSDs).

Lustre implements a loadable Linux kernel module installed
	in every computing node for POSIX compliance. 
The computing nodes are diskless:
	all I/O operations go by RPC to the LNET nodes, and the LNET nodes control the storage access with direct connections. 
\lnet nodes act as virtual switches connecting two different network fabrics:
a proprietary network connecting compute nodes, and an Infiniband network connecting storage servers.

\storage{} employs the following reliability mechanisms.

\paragraph{\bf High availability (HA)}
\storage{}'s HA features are based on server mirroring and data replication, including:
\begin{itemize}
	\item active-active HA pairs for data servers;
	\item active-passive HA pairs for both metadata and management servers;
	\item active-passive for RAID disk devices (OSDs).
\end{itemize}
When a server failure is detected by the partner server, 
    the partner server in the HA pair kills the failed server, mounts the failed server's OSDs, and takes over the failed server's load (to prevent data inconsistencies due to ``split brains'').

\paragraph{\bf Imperative recovery} 
A fail-over triggers a soft restart on the clients that are maintaining connections
	with the failed server; after that
    the management server updates a status table mounted
  at each client.
With imperative recovery, the management server actively informs the clients 
    about the failure to force the
	clients to reload the table and reconnect to the target nodes,
	instead of waiting for an RPC timeout.

\paragraph{\bf Transaction-based recovery.}
\storage{} makes the failover process transparent to storage clients by using
	{\it transaction-based recovery.}
Each client maintains a transaction log. If a server fails,
	the clients automatically reconnect to the new
	server and replay
	transactions that were not committed prior to the failure, in order
	to recover the lost state.

\paragraph{\bf RAID.}
\storage{} uses RAID for disk reliability.
The metadata and management servers are equipped with RAID 1+0 disk volumes,
	and the data servers are equipped with multiple RAID 6 volumes (for storing data blocks)
	and one RAID 1 volume shared between the HA pairs
	(for journaling and state maintenance).
Each volume has two or more hot spares.

\subsection{Failure Characteristics}
\label{sec:study}

Next, we provide characterization of 
storage-related failures, which include 
storage component failures and I/O request 
failures, observed in \storage{} in 23 months between  Jan. 1, 2017 and Nov. 30, 2018.

\subsubsection{Component Failures}
\label{sec:component_failure}

\storage{}-related failures cause service disruption and unavailability. 
For example, in 2018, such failures accounted for 64.4\% (1,175,082 node-hours) of total lost node-hours. 
The total lost node-hours in 2018 was {\em only} 0.74\% (1,825,870 node-hours) of total possible operational hours (of 246,369,160). 
0.74\% is substantial for large-scale data center due to loss of large amounts of compute time; 
in this case, roughly contributing to 32 million core-hours. 
\cref{tab:unavailability} gives a fine-grained 
categorization of the storage-related failures in three different failure domains, as follows.

\paragraph{\bf Client.}
An I/O failure can be caused by failures of a \lustre client module. 
A typical failure mode of a \lustre client module is to hang or crash because of software bugs. Software bugs 
can impact 
	multiple clients at the same time.
In the past, a bug in the lock management
	running on the clients has led to simultaneous failures of hundreds of clients.

\paragraph{\bf Networks.} Network failures lead to unreachability of storage server; thus leading to I/O failures. We categorize
network failures into three sub-domains: 
1) failures of compute-side networks that connect compute nodes, 
2) failures of storage networks that connect storage servers, 
and 3) failures of LNET nodes. 
There were 262, 7, and 6 failures in those categories respectively.
In \storage{}, failures of a network component (e.g., switches)
	require updating routing paths.

	\begin{table}[]
		\centering
		\small
		\begin{threeparttable}
			\begin{tabular}{lp{4.9cm}c}
				\toprule
				{\bf Domain} & \textbf{Failed component} & \textbf{\# Incidents}\\
				\midrule
				Client & Lustre client daemons  & 74  \\
				Network & Compute-side network links/switches & 262  \\
				Network & \lnet nodes  & 6 \\
			 	Network & Storage ToR switch and links  & 7 \\
				Storage & OS/Software  & 11 \\
				Storage & Server HW (CPU/Memory/Fan/PSU/...) &  17 \\
				Storage & Disk drive failures  & 295  \\
				\bottomrule
			\end{tabular}
		\small
		\end{threeparttable}
		\caption{\small Component failures that affected application I/O incidents. Only ten of them caused 
		    system-wide outages (five network-wide outages and five storage-wide outages).}
		\label{tab:unavailability}
		\vspace{-0.4cm}
	\end{table}

\paragraph{\bf Storage servers.}
We find that failures in the domain of the storage servers tend to be more severe and long-lasting,
than client/network failures.
 In total, fail-stop failures led to five system-wide outages
 and multiple partial outages that affected a subset of applications. 
Although most of the storage server failures were handled through the HA features,
 we found that certain fail-over procedures took significantly more time to fail-over
 than usual (several minutes and even hours),
 leading to partial or complete unavailability of the storage system.
 (two such procedures are described in \S\ref{sec:cases}.)  
\storage{} experienced 295 disk failures. Most of them were handled by RAID; however, in 6 of 295 cases,
 disk-drive failures triggered software bugs (5 cases) or there was 1 RAID array failure. In those 6 cases, \storage{} experienced partial system outages:
 hundreds of applications could not connect to the storage system during the failure.

\noindent
\framebox[\columnwidth]{
  \parbox{0.97\columnwidth}{
     Only a very small percentage (0.057\%) of component failures 
        cause system-wide outages. The vast majority of component 
        failures lead to partial system outages or performance issues.
  }
}

\subsection{I/O Failures}
\label{sec:io_study}
We define an {\it I/O failure} as a failure of an I/O request to be serviced
    in the expected time (according to service-level agreements).
From the application's point of view, a late I/O response is 
    no different from a failed I/O response---both of them 
    cause the application I/O to timeout.
In \storage{}, I/O requests are expected to complete within one second.

\subsubsection{\bf I/O Failures Caused by Component Failures}
\label{sss:io_cf}

We find that the most common manifestation of component failures is performance
degradation that leads to I/O failures. 
For example, disk failure is transparently tolerated by the RAID array;
however, disk failures trigger RAID resyncs on hot-spare disks to protect the RAID array from future failures.
Such a resync or periodic scrubbing of a RAID array takes away a certain amount of bandwidth for an extended
period of time, ranging from 4 to 12 hours, and that leads to an increase in completion time of I/O requests. 

\begin{figure*}
\begin{minipage}[t]{0.32\textwidth}
	\includegraphics[]{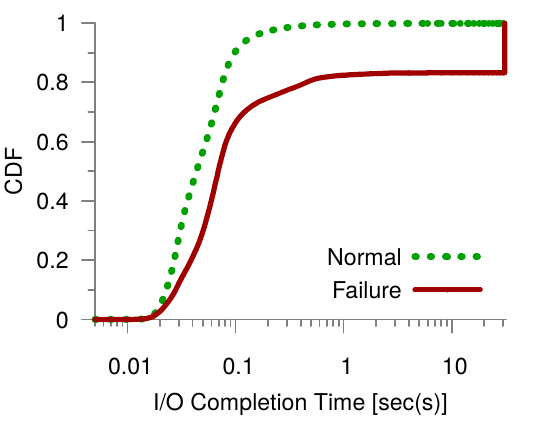}
	\caption{CDF of I/O request completion time under component failures (``Faliure'') and no failures (``Normal'').}
	\label{fig:error_qos}
\end{minipage}\hfill
\begin{minipage}[t]{0.32\textwidth}
  \includegraphics[]{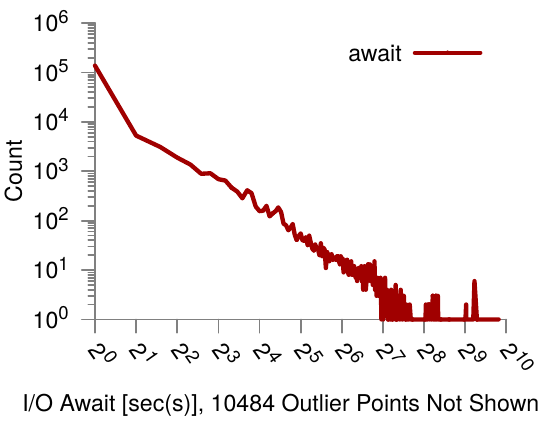}
  \caption{Histogram of disk service time as a load metric. The figure shows that overload 
    is frequent in \storage{}.}
  \label{fig:await}
\end{minipage}\hfill
\begin{minipage}[t]{0.32\textwidth}
	\includegraphics[]{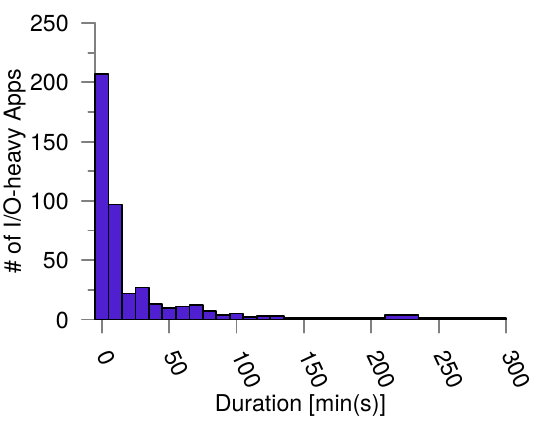}
	\caption{Histogram of duration of high I/O requests per application on a metadata server.
	The tail shows extreme I/O.}
	\label{fig:abusive_apps}
\end{minipage}
\end{figure*}

Using \sysname{}, we find that I/O requests during fail-stop component failures (those that do not
lead to outages) increase the average completion time of I/O requests by as much as $52.7\times$ compared to the average I/O completion time in failure-free scenarios.
Similarly, the 99{th} percentile of completion times of I/O requests is as high as 31 seconds (I/O requests
timeout) due to fail-stop failures. 
Figure~\ref{fig:error_qos} shows the difference between I/O request completion times 
    under component failures and without failures (labeled as ``normal'').
    
\noindent
\framebox[\columnwidth]{
  \parbox{0.97\columnwidth}{
     Component failures have significant performance impact in terms of I/O completion time.
  }
}

\subsubsection{I/O Failures Caused by Resource Overloads}
\label{sss:io_ro}

We find that resource overloads are a frequent root cause of I/O failures.
For a quantitative study, we used {\it disk service time}, 
    returned by \texttt{iostat} as \texttt{await}, 
    as a measure of the load on disk devices.
\texttt{await} measures the average time from the beginning to the end of requests, including device queuing and the actual time to service the I/O request on the disk device.
\texttt{await} is different from I/O completion time, which includes the traversal time between the client and the disk device.
Note that anomalies in disk service time could have different causes,
	such as disk errors, extreme I/O requests that
	content for blocks on a specific disk,
	and high load that exhausts hyper-threads.

\cref{fig:await} shows a histogram of disk service time (\texttt{await}) in seconds returned
	by \texttt{iostat} using an event-driven measurement (which is triggered
	only when \texttt{loadavg}~\cite{loadavg} exceeds 50, to avoid intrusive behavior).
We can see from \cref{fig:await} that such anomalies occur frequently.
Specifically, we found 14,081 such unique events by 
clustering the per-disk continuous data points in time with \emph{service times longer than
	1 second}.

\paragraph{\bf Extreme IO.} 
Extremely high number of I/O requests create high load on the server and lead to high disk contention. Such unintentional extreme usage of storage systems cause performance and stability problems.  \cref{fig:abusive_apps} shows a histogram of the durations of extreme I/O requests by applications to the indexing server.
The durations of high I/O requests are generally small (lasting less than 10 seconds); however, there is a long tail of applications that send high I/O requests for hours, as seen in Figure~\ref{fig:abusive_apps}. For example, in one case, an application caused high load on the metadata server.
The application opened and closed 75,479,396,602 files
in a span of 4 hours, and issued 20,000 I/O requests per second.
During that time, the \texttt{loadavg} increased from 60 to as high as 350. The 50th and 99th percentile
durations of extreme I/O were found to be 12 and 227 minutes, respectively.

\begin{figure}
	\includegraphics{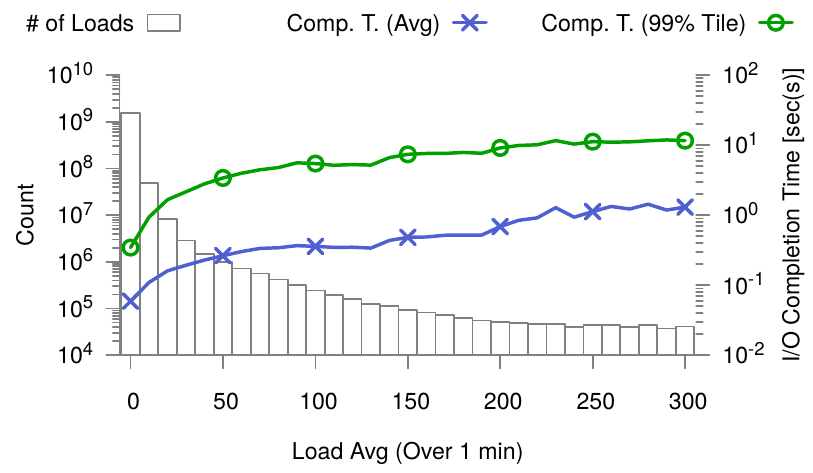}
	\vspace{-0.3cm}
	\caption{Correlation between load (measured by \texttt{loadavg}) and latency. (``Comp. T.'' is the completion time of I/O requests.)}
	\label{fig:high_load}
\end{figure}

\paragraph{\bf High load.} I/O request completion time increases with the load on the storage servers.  High load conditions are caused by a flood of I/O 
requests on a storage server either 1) by one application (cf.~extreme IO), or 2) by multiple
applications that are competing for a shared resource.  Figure~\ref{fig:high_load} shows the histogram
of load across all servers; this graph also shows the average completion time and the 99th percentile completion time of I/O requests (i.e., latency) at different storage server load values. 
Overall, we can see a strong relationship between an increase in load and the completion time of I/O requests.
At high load (\texttt{loadavg} of 350), the average I/O request completion time increases to 1 second, and the 99th 
percentile I/O request completion time increases to 10 seconds. The mean I/O completion time increased $7\times$ under high load (i.e., \texttt{loadavg} $>64$).

\noindent
\framebox[\columnwidth]{
  \parbox{0.97\columnwidth}{
     Resource overloads (due to extreme I/O behavior or high load) are a frequent root cause of I/O failures.
  }
}

\subsection{Long Tail of Zero-day Failures} 
\label{sec:cases}

At the \storage{} scale, failures with recurring patterns have
	all been addressed by building specific solutions over time.
For example, imperative recovery (\S\ref{sec:petasore}) has been deployed
	in response to slow recovery and timeout of I/O transactions,
	and preemptive disk replacement has been adopted to prevent concurrent
	disk failures and reduce recovery time.
On the other hand, we have found a long tail of failures that
	do not have common patterns and are hard to anticipate; we
	call them {\it zero-day failures} (analogous to
		zero-day vulnerabilities)~\cite{Mogul:2017}. 
		
We analyzed issue tickets and found that there were four new issues on average per month.
The following describes two cases of zero-day failures, which by definition are rare.

\paragraph{\bf Failures of failure recovery.}
\storage{} has experienced multiple failures of failure recovery throughout its lifetime. 
The failures inevitably led to unavailability of the largest partition on \storage{} for several hours.
The management server (\mgs) detected that the active metadata server (MDS-A) was not
responding because of a software bug that had led to a hang.
Ten minutes later, after several retries \mgs declared the MDS-A as unreachable.
22 minutes after \mgs had detected a problem in the MDS-A, and launched the fail-over procedure 
    by which the active standby MDS-B mounted MDS-A volumes and triggered imperative recovery.
During the mounting process, the RAID array was found to have errors which
	triggered a background reconstruction process that used a hot spare.
Because of the RAID reconstruction, the recovery of client transactions timed out,
causing client eviction. 150 minutes later, the RAID array reconstruction, combined
with high I/O requests, overloaded MDS-B and caused the whole storage partition
to become unavailable. 
During the 4 hours of this case,
the whole file system partition provided a degraded service for
about an hour and was unreachable for the remaining 3 hours.
During that outage, jobs running on 8,346 compute nodes were interrupted.

\paragraph{\bf Failures of \lnet nodes.}
\lnet nodes serve as bridge devices between computing nodes and storage servers.
A request from a client to an \ost (a RAID disk device) can be served by any one of 4 \lnet nodes.
For any pair of <client, OSD>, the group of 4 \lnet nodes are fixed and chosen in round robin when routing a request. 
A failure of an \lnet node is detected through heartbeats.
However, in this case, an \lnet node was found to drop requests passing through it, causing I/O failures. 
The \lnet node appeared to be alive and healthy and sending heartbeats to the rest of the components.
Upon investigation, it was found that the \lnet had suffered a software error
that caused it to drop I/O requests.
The incident was captured via an application's own performance monitoring system.
The I/O bandwidth (in MB/sec) for the applications served by the failed \lnet node decreased by 25+\% for multiple hours.

\noindent
\framebox[\columnwidth]{
  \parbox{0.96\columnwidth}{
     Many high-impact zero-day failures can be prevented if the faulty 
     or unhealthy components can be detected and the corresponding potential causes 
     can be diagnosed earlier, before 
     they lead to user-visible impact.
  }
}

\section{Design and Implementation}
\label{sec:design}

\begin{figure}
	\includegraphics[width=0.42\textwidth]{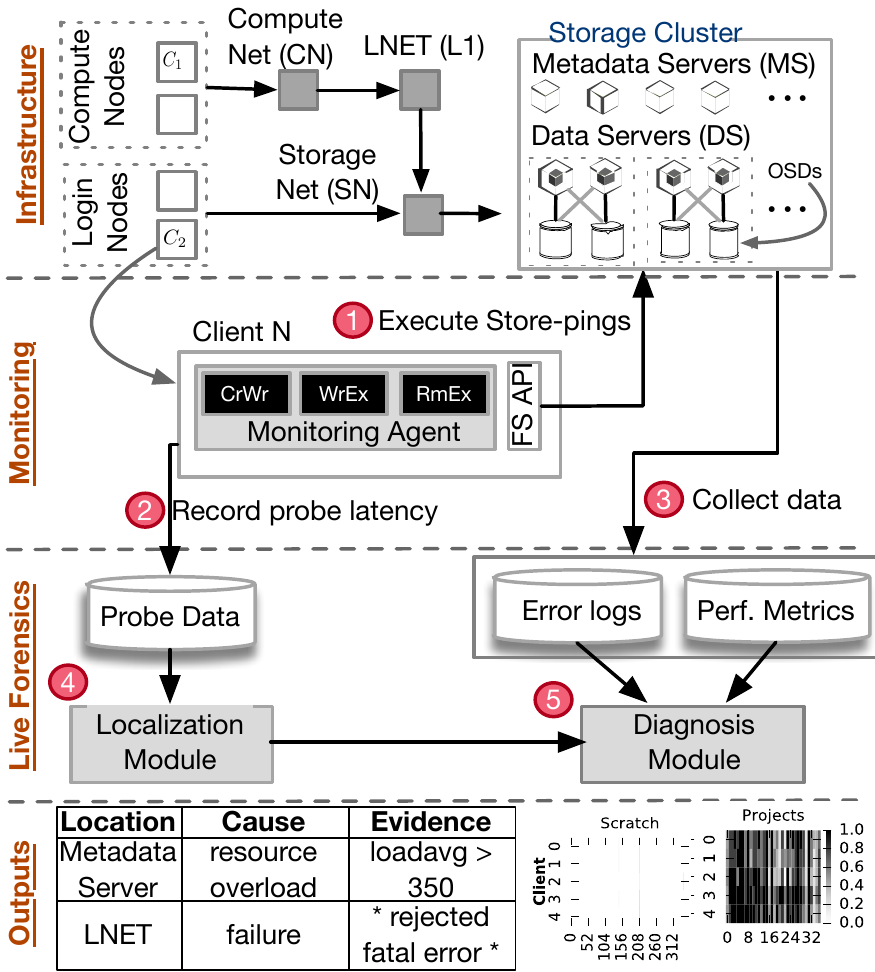}
	\caption{An overview of \sysname. \sysname{} consists three component for monitoring,
	    failure localization, and failure diagnosis (marked in gray).}
	\label{fig:klscope}
    \vspace{-0.3cm}
\end{figure}

Figure~\ref{fig:klscope} shows the overall architecture of \sysname{} and how it 
    fits to a large-scale distributed storage system like \storage{}.
\sysname{} has three main components: 1) proactive monitoring modules 
    for failure detection,
    2) failure localization based on modeling of the health of every storage component
        (e.g., metadata server and data server)
        using the monitoring data,
and 3) a diagnosis module that pinpoints the root causes of unhealthy storage 
    components that affect the performance of I/O requests.
    
\sysname{} is a fully automated system. 
Its design follows the principle in \S\ref{sec:principle}.
A key emphasis of its design is the need to
    be practical and useful for real-world production
    storage systems like \storage{}.
Its success has been proven by its deployment at \storage{}
since 2017. \sysname{} has been used 
    by \storage{} operators to quickly identify failing components, perform preemptive replacement, and help users rewrite application code to avoid extreme I/O.

\subsection{Monitoring}
\label{sec:monitoring}

\sysname{} {\it proactively} monitors a storage system from the viewpoint of both the clients and the storage components.
The component view alone
	is insufficient to reveal and identify partial,
	gray failures manifested in complex performance anomalies,
while the client view alone is insufficient
	to reveal the root causes inside the storage subsystem.

\subsubsection{Store-Pings}
\label{sec:store_ping}
We designed {\it Store-Pings}, a set of primitives for measuring the availability
	and I/O performance
	of distributed storage systems.
Store-Pings are analogous to but different from the ICMP-based network \texttt{ping}.
A Store-Ping is a specific I/O request that
	traverses all necessary storage components to complete the I/O request.
The API of a Store-Ping is:
\begin{center}
\texttt{store\_ping(ost, *io\_op, kwargs)}
\end{center}
where \texttt{*io\_op} is a function pointer to an I/O operation,
and \texttt{kwargs} is the argument for 
\texttt{*io\_op}. Store-Pings use direct I/O 
requests to avoid any caching effect into the client 
memory, thus ensuring the I/O request's traversal of the disks on the data servers.

We designed three types of Store-Pings, \texttt{CrWr}, \texttt{WrEr}, and  \texttt{RmEx}, which
	represent three different I/O requests:
\begin{itemize}
	\item \texttt{CrWr}: create and write a file
	\item \texttt{WrEx}: write to an existing file
	\item \texttt{RmEx}: remove an existing file
\end{itemize}
The three types of Store-Pings test all the storage subsystems
	involved in ensuring correct and successful I/O operations.
\texttt{CrWr} and \texttt{RmEx} 
	test the functionality of the metadata servers, whereas \texttt{WrEx} tests the functionality of the 
	data servers (and, correspondingly, RAID disks).
For example, a \texttt{CrWr} requires two different backend operations to complete:
	(1) creation of a file by a metadata server on a random data server (and the corresponding RAID
	disks) and adding of the file entry to the metadata index,
	and (2) opening and writing of a file on the data server (and the corresponding RAID disks).
The payload of a write request is as small as 64 bytes.

Store-Pings assume that a {\it monitor} can pin files onto a specific
OSD (and hence the data server). File pinning can be easily 
supported, as the metadata server
	has all the data chunk information.
Thus, a file uniquely identifies a data server for a monitor, and that also helps
	to minimize variation in latency measurements (avoiding
	concurrent Store-Pings from other monitors).
	This functionality is used to deterministically test the data servers by using 
	\texttt{WrEx} 
	I/O requests. Store-Pings do not require pinning of any other nodes (e.g., metadata servers).

\subsubsection{Store-Ping-Based Monitors}

A monitor continuously executes 
Store-Pings to measure the availability and 
performance of storage components. 
However, in order to monitor all 
storage components (LNETs, metadata servers, data servers, and OSDs), \sysname 
needs to select the number of 
Store-Ping monitors and their 
placement in the compute network.  Store-Ping monitors should be enabled only on a subset of clients ($M$ out of $N$ clients where $M<<N$) to reduce the overhead of the monitoring system and its impact on existing I/O requests. Thus, selection of the number of Store-Ping monitors, $K$, and their placements can be formulated as a constraint
	optimization problem: the monitors should achieve
	the highest coverage (observability)
	under the monitoring overhead 
	budget. 
	
	In this work, we use network tomography principles~\cite{castro2004network} in which the aim is to find unhealthy components in a given network topology graph $G$ by running tests on subsets of components, with each test (which is a completion of an I/O request sent by Store-Pings) indicating whether any component in the subset is unhealthy. The subsets of components that can be tested together are limited by the set of measurement paths $P$, which are in turn limited by the topology, probing mechanism, and placement of monitors. Specifically, the placement of monitors 
	in \sysname{} is guided by the ``sufficient identifiability condition''~\cite{Ma:2014:NFL:2663716.2663723}, which states that in a 
	network graph $G$ consisting of both 
	monitor and non-monitor nodes, 
	any set of up to $k$ failed nodes 
	is identifiable if for any 
	non-monitor $v \in G$ and failure 
	set $F$ with $|F|\le k$ ($v 
	\not\in F$), there is a 
	measurement path going through 
	$v$ but no node in $F$. Such a 
	method ensures {\it spatial 
	differential observability}. 
	These principles put a restriction on the minimum number of concurrent failures (in our case, unhealthy components), $k$, that can be detected in the storage system.%
	
\subsubsection{Component-Side Monitoring}
\label{sec:comp_monitor}
Store-Pings can be used to infer the health of storage components based on client-view observations;
however, it does not provide any information about the root cause of unhealthy components. 
\sysname{} uses a comprehensive component monitoring system (Integrated System Console \cite{6968671}) to collect 
    status data
	on different components, including I/O statistics (\texttt{loadavg} and I/O requests per second 
	of data servers, and
		I/O wait time and  utilization of disks) and
		logs (error logs and syslogs) of each storage server.

\subsection{Failure Localization}
\label{sec:triage}

\sysname{} infers the health state of a storage component based on Store-Ping monitoring. 
It addresses the challenges of
1) {\it measurement noises} due to asynchronous and variability in I/O measurements,
and 2) {\it non-determinism} due to path redundancy and randomness in routing/balancing.
Our experience tells us that threshold or voting-based methods are ineffective,
    because
    of the non-linearity that originates in measurement noise and non-determinism.

\sysname{} models health as a stochastic process that accounts for 
    measurement noise and non-determinism. 
The modeling is based on factor graphs (FG)~\cite{koller2009probabilistic}, a generalization 
of probabilistic graphical models. %
FGs model the relationship between component health (random variables) 
using domain-guided {\it factor functions} (how a component's health leads to I/O failures). 
Factor functions encode 1) the stochastic nature of measurement noise and variability, 
and 2) non-determinism due to path redundancy. 
FGs only needs small samples and 
    outperforms supervised machine learning (e.g., regression~\cite{Zhang:2018,tan2019netbouncer})
    with domain-guided factor functions.

In this section, we formalize the modeling and explain the procedure of 
    using the model to infer the health of the components, including metadata servers,
    data servers, I/O forwarding servers (LNETs), and the storage network.

\paragraph{\bf Modeling.}
We define the {\it health} of a component as the probability that it will successfully serve an I/O request,
    denoted by $X_i$. 
$X_i$ is sampled from a beta distribution, $X_i \sim Beta(\alpha, \beta)$, to incorporate uncertainty,
    where $\alpha$ and $\beta$ determine the shape of the distribution.\footnote{Beta distributions are continuous distribution commonly used as a prior for Bernoulli random variables.
    It drastically reduces the computation time for inference.}
At any epoch (i.e., at the time of inference), $\alpha$ and $\beta$ are updated based on historical information up to the current epoch.

Store-Ping-based monitoring provides measurements on path availability between the two members of each <client, OSD> pair. 
We use the random variable $Y_{p}$ to denote the number of observed successful Store-Pings on a path $p$ (a <client, OSD> pair), 
and we model $Y_{p}$ using a binomial distribution, i.e., $Y_{p} \sim Binomial ({A}_{p}, N)$, 
    where ${A}_{p}$ denotes the availability of the path $p$ and $N$ denotes the number of Store-Pings sent by the 
    monitoring system through $p$. 

We leverage the fact that, for a Store-Ping to be successful, 
every component on the I/O path must be both
available and reachable from the client. When
a Store-Ping takes a unique path from a client to a \ost,  
the path availability ${A}_{p}$ can be determined solely by the product of individual component's health: 
    ${A}_{p} = \prod_{i \in \mathrm{p}}{X_i}$.
    
\begin{figure}
\centering
\includegraphics[width=0.4\textwidth]{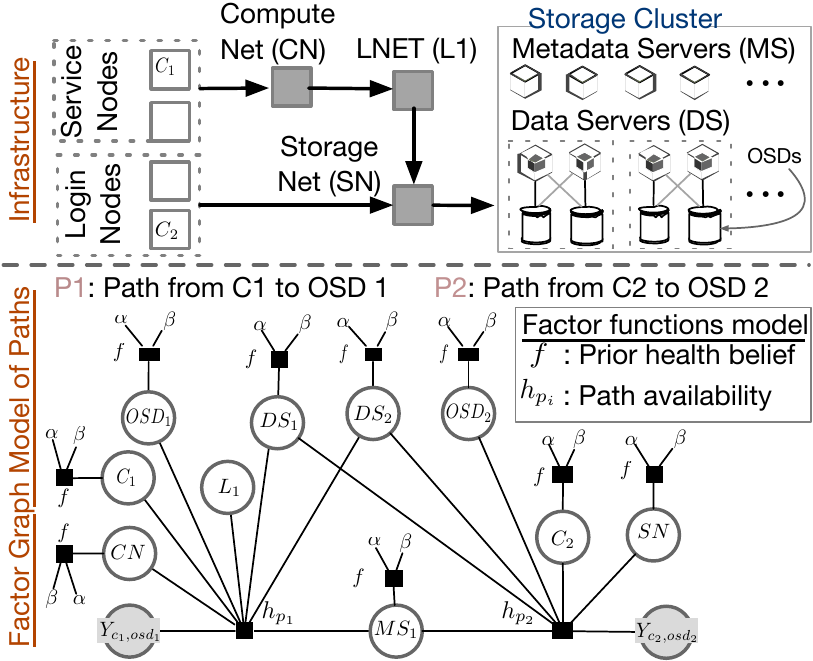}
\caption{An illustration of the FG model. Only the paths $C_1$ to $OSD_1$, and $C_2$ to $OSD_2$ are shown, for clarity. Redundancies and other network components have also been removed for clarity.}\hfill
\label{fig:hcfg}
\end{figure}

However, because of the redundancy in a distributed system 
(e.g., the HA pair-based failover), a Store-Ping destined for an \ost may take a different path. 
Thus, the path availability cannot be expressed in terms of each component's own availability, but also 
depend on the availability of the other redundant 
components. For example, an I/O request to an \ost can be 
routed through one of two data servers connected to it.
Hence, a destination \ost is not reachable 
if both of the two data servers connecting to it are not 
available or the \ost itself is not 
available. $R_{osd_{i}}$, the probability of reachability of 
the \ost, can be 
determined as:
\begin{center}
$R_{osd_{i}} = (1 - (1 - X_{ds_1}) \cdot (1 - X_{ds_2})) \cdot X_{osd_i}$
\end{center}
where $X_{ds_1}$ and $X_{ds_2}$ denote the health of data servers in the HA pair 
associated with the OSD (denoted by $osd_i$). In this 
equation, $1-X_{ds_1}$ and $1-X_{ds_2}$ determine the 
probability distribution of the $ds_1$ and $ds_2$ to be {\em 
not healthy} respectively, and their product determines the probability 
distribution of both being in unhealthy state. 
The probability distribution, when multiplied by the 
probability distribution of the \ost is health, gives the 
reachability of the \ost. From \cref{fig:hcfg},  $A_{P}$ between client $C_1$ and $\ost_1$ is given by:
\begin{center}
    $A_{P} = X_{C_1}\cdot X_{CN}\cdot X_{SN} \cdot X_{MS1} \cdot R_{osd_{1}}$
\end{center}
Thus, the path availability $A_{P}$ must explicitly model such redundancies  (e.g., LNETs and HA-pairs) while estimating the availability of a path.

The model described above can be represented using a factor graph 
that models the interactions between different random variables (shown as circles) and 
{\it functional relationships} known as Factor Functions (shown as dark boxes).

Figure~\ref{fig:hcfg} 
shows a part of the FG that models 1) the health of components  that lie on the 
path of <$C_1$, $\ost_1$> and <$C_2$, $\ost_2$>, 2) path availability for these components, and 3) \ost availability. The 
components $OSD_1$, $OSD_2$, $DS_1$, and $DS_2$ form an HA 
group (\S\ref{sec:petasore}). The circles in the FG represent random variables (e.g., a component's health). 
The factor functions, represented by squares, encapsulate the relationships among the random variables. 
The singleton factor functions $f_i$ encapsulates the prior belief of the health of the component, which is given by the Beta distribution (see above). The multivariate 
factor function $h$ models the number of successful Store-Pings on a path, which is given by the binomial distribution (see above).

\paragraph{\bf Inference.} With HCFG, we can calculate the health of each component $X_i$ in the system.
The expected health of a component $i$ can be estimated as $E[X_i|Y_{p_1}, Y_{p_2}, Y_{p_3},...]$. 
Observations ($Y_{p_1}, Y_{p_2}, Y_{p_3},...$) 
and the prior belief of the health of components ($\alpha$ and $\beta$ for each $X_i$) are needed 
at epoch $T_j$. $Y_{p_i}$ is measured by the number of observed successful Store-Pings during a specified interval, 
and $\alpha$ and $\beta$ are obtained from the inference result at the previous epoch. 

We solve the inference task using the Monte Carlo Markov Chain (MCMC) algorithm~\cite{neal1993probabilistic}. MCMC is a technique that can be used to estimate the expectation of a statistic from a complex distribution (in this case, $E[X_i|Y_{P_1}, Y_{P_2}, Y_{P_3},...]$)  by generating a large number of samples from the model and directly estimating the statistic.

\subsection{Failure Diagnosis}
\label{sec:rca}

After identifying an unhealthy component, \sysname{} further pinpoints the
 causes, including  component failures (\cref{sss:io_cf}) or
	resource overloads (\cref{sss:io_ro}).
\sysname{} pinpoints the corresponding error logs for component failures, 
    and the corresponding load metrics for resource overloads.

The root cause analysis is based on a statistical method.
Our key insight is that unhealthy components and their healthy counterparts behave differently.
For example, we expect all data servers that belong to the same file system
	to generate similar error logs over a time window.
However, when a server is unhealthy that group 
	it generates unique patterns of error logs.

\paragraph{\bf Attributing the cause to overloads.}
We collect the following monitoring data (\S\ref{sec:comp_monitor})
to identify resource overloads as the cause:
\begin{itemize}[noitemsep,nolistsep,leftmargin=*] 
	\item \emph{loadavg} on the data and metadata servers which captures the load on a server at 5-minute intervals;
	\item \emph{await} time of disk devices, which captures the average service time (in milliseconds) taken by a disk device to serve an I/O request; and
	\item \emph{utilization} of disk devices, which captures the bandwidth saturation of a disk device.
\end{itemize}
\sysname runs a local outlier factor (LOF) anomaly detection algorithm~\cite{Breunig:2000:LID:335191.335388}
on a homogeneous group of storage components. The LOF is based on the concept of
a local density, where locality is given by k-nearest neighbors, whose distance
is used to estimate the density.
By comparing the local density of an object to the local densities of its neighbors,
one can identify regions of similar density, and points that have a substantially
lower density than their neighbors, and thus are considered to be outliers.
We chose LOF because storage components within a homogeneous group may have
different modes of operation. Such modes are not indicative of any degradation.
For example, in the production data, we found cases when $k$ data servers had
low \texttt{loadavg} (<10) and $N-k$ data servers
had high \texttt{loadavg} (>64). However, if there is a data server whose
\texttt{loadavg} is significantly higher than both of those  modes of operation,
that is indicative of the problem.
A similar methodology is adopted for disk devices by using the await and
utilization metric.

\paragraph{\bf Attributing the causes to component failures.}
Component failures are attributed based on log analysis.
\sysname{} collects error logs from all the components of the storage subsystem
(\S\ref{sec:comp_monitor}).
The error logs generated by the unhealthy components are compared to
	the error logs of healthy components, $\delta = L_{\text{UO}} -\bigcup\limits_{i \in HO} L_i $,
	where $L$ represents the log set,
	$UO$ represents unhealthy components, and $HO$  represents healthy components.
If $\delta \ne \varnothing $, then $\delta$ is provided as evidence, and the unhealthy status
	is attributed to component failures.
Note that \sysname{} processes the raw logs and curates them into a normalized
	form that captures the triggering events of the logs by filtering out
	time- and node-specific metadata.
It does so using regular expression based log processing tools~\cite{taerat2011baler} based on filters  
provided by
	production facilities
	for the Lustre file system.
Despite its simplicity, we find that the statistics-based log analysis is very
	effective and scalable in pinpointing the log entries
	that indicate the failure causes,
	as shown in our evaluation.

Note that \sysname could report the likely causes of IO failures
	to be marked as both component failures (by logs)
	and resource overloads (by metrics).
	
\subsection{Implementation and Deployment}
\sysname{} has been deployed in 
\storage{}. 
We placed monitors on clients that
	(1) have different underlying system stacks (e.g., kernel versions),
	(2) are physically located on different networks, and
	(3) execute different services (e.g., scheduling, user login, and data moving).
Specifically, we placed monitors on all the service nodes (64 nodes) that provide scheduling and 
other services,
	import/export (I/E) nodes that help move bulk data into and out of the storage system (25 nodes), 
	and login nodes (4 nodes) that users use to launch applications. The I/E nodes and login nodes are 
	located on the storage network, whereas the service nodes are located on the proprietary compute network fabric.  However, in  production, at any given time, Store-Pings are executed from (1) all login 
	nodes, (2) 1 out of 64 service nodes chosen randomly, and (3) 1 out of 25 I/E nodes\footnote{I/E nodes are the import export nodes that are used to move data in and out of the \storage{}.} chosen 
	randomly. This probing plan not only satisfies our minimal probing plan for inferring storage system 
	health, but also provides reliability of monitoring infrastructure itself. That is, in case of a client-failure another client can be chosen as a monitor. 
	
Store-pings are executed every minute for each OSD, data 
	server and metadata server. For data and metadata servers, the Store-Pings use the same \lustre APIs used for writing to disk, but instead create/read/write to the memory of the server. 
	This results in 72 \texttt{CrWr} (6 
	clients $\times$ 6 metadata servers and 6 metadata OSDs), 72  
	\texttt{RmEx} (6 clients $\times$ 6 metadata servers and 6 metadata OSDs) and 5,184 \texttt{WrEx} (6 clients $\times$ 
	432 data servers and 432 OSDs) requests/minute.
	To get deterministic measurement paths for Store-Pings between the two members of each $<$monitor, OSD$>$
	pair, we use \texttt{setstripe} of Lustre to create a unique file on each OSD for a client.
	
The Store-Ping-based monitoring is implemented in Python 
and scheduled using Jenkins~\cite{jenkins}. 
The Store-Pings are configured to run at one minute intervals with a timeout of 30 seconds for each I/O request. 
The failure localization module is implemented using PyMC3~\cite{10.7717/peerj-cs.55}, 
    a Python-based probabilistic programming language. The failure localization module uses samples collected over five minutes; thus, it uses data from 26,640 I/O requests for inference. 
Finally, the diagnosis methodology was implemented in Python using the Scipy~\cite{jones2014scipy} and Baler~\cite{taerat2011baler} libraries.
\section{Evaluation}
\label{sec:eval}
We evaluated \sysname{} using 843 production issues resolved by the
	\storage{} operators over a span of two years (Dec. 1, 2016 to Nov. 30, 2018.).
Each of the 843 issues has a
	report with manual categorizations
	that we refer to as the {\it ground truth}.
As \sysname{} has been deployed for more than two years,
	we used \sysname{} to do live forensics (triage and root-cause analysis)
	for each of the issues.
As discussed in \S\ref{sec:eval:overall}, \sysname{} reports far more
	issues than the 843 found but it is up to the operators to decide whether to
	investigate a particular issue, based on its severity, job priority,
	and communications with customers.

\subsection{Overall Results}
\label{sec:eval:overall}
Table~\ref{tab:effectiveness} presents the effectiveness of \sysname{}
	in triaging the failing components (\S\ref{sec:triage})
	and pinpointing their root causes (\S\ref{sec:rca}).
We can see that \sysname{} can localize the unhealthy components, caused by failures or \overload,
	for 99.3\% of the production issues (837 out of 843).
Only six out of 843 were not detected by \sysname{}.
We find none of the six issues had any impact on the I/O completion time.
All six issues belonged to disk drive failures. Those failures were recorded and flagged for repairs to 
avoid RAID failures.

Among the 843 production issues, 346 were caused by component
	failures and 497 were caused by resource overload, as discussed in \S\ref{sss:io_cf} and \S\ref{sss:io_ro},
	respectively.
As shown in Table~\ref{tab:effectiveness}, 
\sysname was able to correctly identify the root causes of 98.3\% of the issues caused by component failures. 
In addition to correctly associating the root cause of
and issue to a failure, it presented system managers with
 the error logs corresponding to the failure.
For overload issues, \sysname{} was able to correctly associate the
root causes of 94.2\% of the issues (468 of 497)
while incorrectly attributing the remaining 29 issues to component failures.
The reason is that the 29 overload issues coincidently had
	random noises in the logs, which confused \sysname{}.

\begin{figure}
	\includegraphics{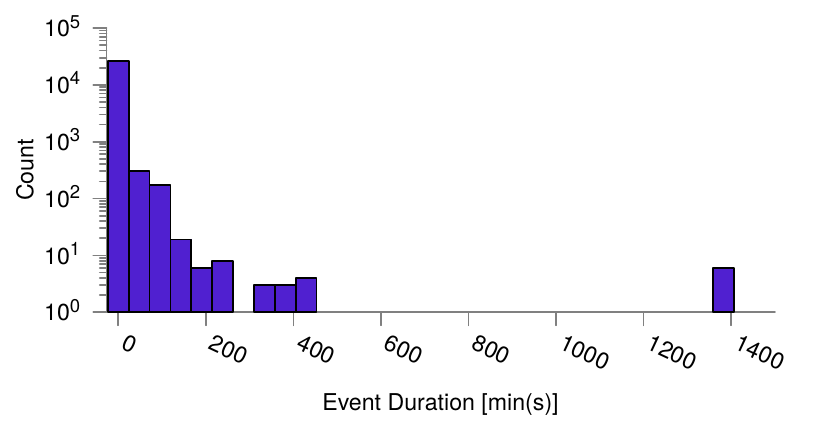}
	\caption{Histogram of duration of  components in unhealthy state.}
	\label{fig:hist_kscope_localization}
\end{figure}

In addition to the 843 known issues, \sysname found another 25,753 I/O failure events.
Figure~\ref{fig:hist_kscope_localization} shows the histogram of
 	the durations of these I/O failures.
As shown in the figure, 6073 of 26,596 I/O failures lasted for more than 5 minutes;
	1,773 lasted for more than 20 minutes;
	and 1,026 lasted for 30+ minutes.
We find that the 843 reported cases mostly fall into the range of
	20--30 minutes.
Typically, operators focus only on issues that last for 30 minutes.

\begin{figure}
	\includegraphics{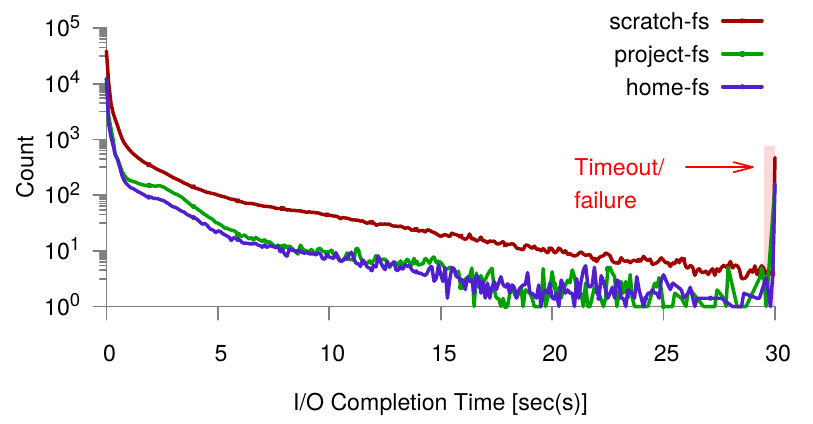}
	\caption{Completion time of I/O requests measured by \sysname's \texttt{WrEx} Store-Pings.}
	\label{fig:eval:latency}
\end{figure}

Our interactions with \storage{}'s operators told us that
	the Store-Ping-based monitoring helped them understand
	the tail latency and performance variation in real time.
Operators can detect performance regression by comparing the measurements
	from different points of time.
Figure~\ref{fig:eval:latency} shows the latency measurement histogram
	(plotted as a line with every 20 points on the graph) for the \texttt{WrEx}
	Store-Pings. (We omit \texttt{RmEx} and \texttt{CrWr} because of the page limit.)
We can see that 99\%  of \texttt{WrEx} completed within one second (SLO),
	and only 0.14\% failed with a timeout.
With \sysname{}, it is efficient to nail down to the anomalies and perform
	live forensics (e.g., the load-related resource overload condition discussed in \cref{sec:study}).

\begin{table}[t]
		\begin{tabular}{lccc}
			\toprule
			      & {\bf True Positive} & {\bf False Negative} & {\bf Total} \\
			\bottomrule
		\rowcolor{gray!15}
		\multicolumn{4}{l}{Component Triage (Total: 843)} \\
			      & 837 (99.3\%) & 6 (0.7\%) & 843 \\
		\rowcolor{gray!15}
		\multicolumn{4}{l}{Root-cause Analysis (Total: 843)} \\
		(Failure)  &	340 (98.3\%)  & 6 (1.7\%) &  346 \\
		(Overload) & 468 (94.2\%) & 29 (5.8\%) & 497 \\
			\bottomrule
		\end{tabular}
		\caption{Effectiveness (measured by true positives)
			of \sysname's triage and root-cause analysis.}
		\label{tab:effectiveness}
\vspace{-20pt}
\end{table}

\subsection{\bf False Positives and Incorrect Diagnosis}
\label{sec:fp}

It is challenging to measure the false positives of \sysname{}, as there are
	no ground truth data; an issue tagged by \sysname{} as not being resolved
	could come from different reasons (e.g., low priority jobs).

To estimate the false-positive ratio,
we randomly selected 100 issues reported by \sysname{}: 50 from the
	``component failure'' category and 50 from the ``\overload''
	category.
\sysname was able to localize {\it all} cases correctly. 
However, it failed to identify the root causes of four (out of 100) cases. 
Our further investigation showed that the false positives
were due to the propagation delay between the occurrence of the internal faults at the
	server side
and their manifestation as I/O failures. Such time misalignments are
expected in a production system, and \sysname currently does not model them.

\subsection{Monitoring Overhead}
We use the IOR benchmark~\cite{ior} to measure the monitoring overhead in a worst-case scenario.
The measurement uses stress testing to max out the throughput offered by \storage{}. 
IOR was running on 4,320 compute nodes during this measurement. \cref{tab:impact-qos} shows the monitoring overhead 
introduced by Store-Pings when (a) 100 monitors were running at 30 second interval 
and (b) 6 monitors were running at one-minute interval. Store-Pings decreased mean throughput {\em only} 
by  $<0.01\%$ in \storage{}'s production settings. 
However, scaling to 100 monitors and increasing the frequency by 2$\times$ would 
    decreases the throughput by less than $2.42\%$. 
Note that the average throughput in production is significantly below the peak throughput 
under the stress test. We also measured the time difference between the launch of Store-Pings for a given interval and 
found that all Store-Pings were launched within 10 seconds of each other and 98.4\% were launched within 3 seconds of each other.

\begin{table}
\begin{tabular}{|l|l|l|l|l|}
\hline
      \textbf{\sysname{}}      & \multicolumn{2}{l|}{\textbf{100 monitors}} & \multicolumn{2}{l|}{\textbf{6 monitors}} \\ \hline
             & \textit{Mean}        & \textit{Std}        & \textit{Mean}       & \textit{Std}       \\ \hline
\textbf{Off} & 100                  & 0.15                & 100                 & 0.13               \\ \hline
\textbf{On}  & 97.58                & 0.32                & 99.99               & 0.12               \\ \hline
\end{tabular}

		\captionof{table}{Impact of 100 Store-Ping monitors running at 30 second interval on IOR benchmark~\cite{ior} for stress testing. The mean value of I/O throughput with no \sysname is normalized to 100.}
		\label{tab:impact-qos}
\end{table}

\subsection{Simulation}
\label{sec:simulation}
Before deploying \sysname{} in production, we built a trace-based simulator
	based on \storage's topology
	to extensively evaluate the localization and root cause analysis.
We ran 1,000 simulation experiments. In each experiment, the simulator injected faults based on the distributions of completion
	time characterized in \S\ref{sec:study} to simulate both component failures
	and \overload.

In 1,000 simulation experiments, when the simulator injected exactly two faults (one component failure and one overload), \sysname detected
	these cases with no false positives.
We then increased the number of simultaneous faults to 20 for failures and {\overload}s.
With 40 faults, in the worst case, \sysname generated 4 false positives and found all the injected faults (recall of 1.0).
\sysname{} performed particularly well in the simulations because
 the simulated faults have immediate manifestation and less noise.

\begin{figure*}[t]
\centering
	\begin{subfigure}{3in}
		\includegraphics[]{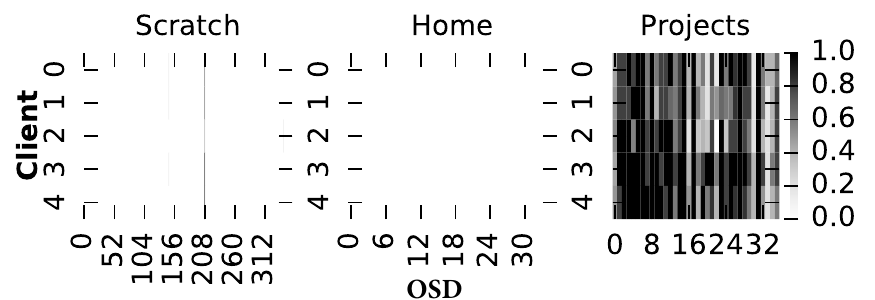}
		\caption{Issue on Scratch OSD 208 and Projects MDS}
		\label{subfig:kl-mds}
	\end{subfigure}
	\begin{subfigure}{3in}
		\includegraphics[]{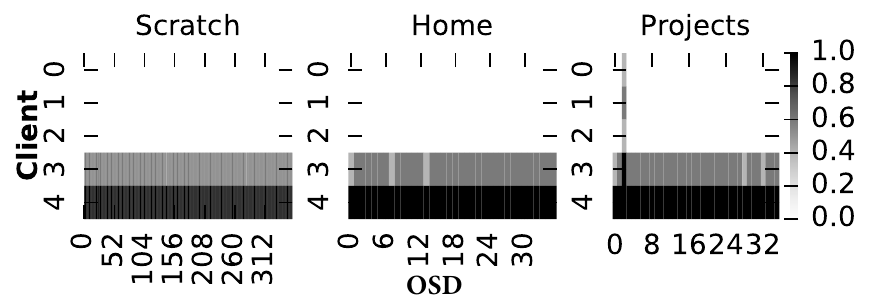}
		\caption{Issue on compute network}
		\label{subfig:kl-net}
	\end{subfigure}
	\vspace{-0.3cm}
	\caption{Outages visible from \sysname}
	\label{fig:kl-examples}
\end{figure*}

\subsection{Kaleidoscope in Action}

We use real I/O failures to illustrate \sysname in action. 
\cref{fig:kl-examples} uses a heatmap to depict a failure impact on data servers. 
Each heatmap shows the ratio of operations that took longer  than 1 second to the total number of operations issued during 5 minutes interval by a given client (y-axis) to each data server from Scratch, Home, and Projects domains (x-axis). Scratch, Home, and Projects are three file systems in \storage{}. Clients 0, 1, and 2 are the login nodes on Ethernet network, client 3 gives an aggregated view of all 25 IE nodes on Infiniband network, and client 4 provides an aggregated view of all 64 service nodes on compute network.

\cref{subfig:kl-mds} shows a case of two failures that occur concurrently in the storage system: (i) a load issue on scratch data server 208 and (ii) an outage of projects file system metadata server.  %
The heat map in \cref{subfig:kl-mds} shows that both scratch and home data servers are healthy for most part with exception of the scratch data server 208. Data server-208 is unhealthy as all clients have trouble completing I/O requests within 1s. The heatmap for projects file system indicates a wide-spread outage that could be caused by network-wide issue, metadata server outage, or concurrent outage on all 36 \lnet nodes connecting clients to projects file system. As \sysname{} shows that scratch and home data servers are functioning properly, network-wide issue is improbable as they all share the same network. It is also highly unlikely that 36 \lnet concurrently fail. Therefore, by elimination, we conclude that the observed outage would only be caused by an issue with: (i) projects file system metadata server or (ii) the 2 LNETs serving the metadata server.

The heatmap in \cref{subfig:kl-net} shows that all requests from aggregated view of client 4 are failing (or took longer than 1s) across all data servers. The ratio of 1.0 across all data servers is a clear indication of a network-wide outage. 
In case of IE nodes (i.e., Client 3), one of the clients  is having trouble accessing the OSDs across all data servers as indicated by the grayish pattern on the heatmap. From the heatmap alone it is unclear whether the issue was caused by a bad \lnet node or the client itself. However, it is possible to diagnose the problem by looking at the topology of the system to rule out contribution of \lnet or client for failing I/O requests. 

In both cases, \sysname{} provides correct localization and diagnosis. 
\sysname provides {\em only} relevant visualizations after localization between OSD and clients 
    as the evidence, instead of generating visualizations for all possible combinations (e.g., \lnet and clients). 

\section{Discussion and Limitations}
\label{sec:discussion}

\sysname{} is designed with an emphasis on practicality and scalability,
    without a desire to be 100\% accurate.
As we show in \S\ref{sec:fp}, \sysname{} occasionally introduces 
    false positives and incorrectly diagnoses
    due to not modeling the propagation delay between the cause and the impact.
We show that the false positive ratio is very low,
    and our interactions with \storage{} operators confirm 
    its usefulness.
    
One potential caveat is that \sysname{} assumes that Store-Pings experiences 
    the same I/O behavior as real applications, which may not hold in all cases.
On the other hand, using \sysname{} ML-components retrospectively on trace-data generated by Store-Ping monitors show that 
    ML-methods captured majority of the failure and resource contention problems (see \cref{sec:eval}).
    
Note that different applications may require different requirements of I/O completion time.
    For example, some applications are more tolerable to slow I/O response.
\sysname{} does not understand application-level semantics but uses a system-wide 
    threshold based on the service-level agreement (one second for \storage{}).
An alternative we considered was an approach that inserts Store-Pings in the I/O path of 
    the applications by automatic binary instrumentation or source code analysis.
However, we find that this is not feasible in data center settings such as \storage{},
    in which we have little control over applications, 
    not to mention the concerns of correctness and performance overhead.

In the deployment at \storage{}, \sysname{} detects errors at the granularity of 
    meta-data servers, data servers, \lnet and object store devices. It cannot detect finer grained faults such as 
    section errors~\cite{Bairavasundaram:2007,Schroeder:2010}.
It currently can only localize a network outage related I/O failure,
    instead of the specific network router, due to the randomness in network routing decisions.
This is a deployment decision rather than a fundamental limitation of the methodology.
To have more insights into the network, one can enhance the probes in \sysname{}
    based on network tomography~\cite{Zhang:2018,Tan:2019,Arzani:2018,Guo:2015,Geng:2019}.
Note that doing this brings benefits over existing network tomography~\cite{Zhang:2018,Tan:2019,Arzani:2018,Guo:2015,Geng:2019}
    as \sysname{} provides more and stronger capability (e.g., disambiguiting 
        loads and failures)
    as discussed in \S\ref{sec:motivation}.

\section{Related Work}
\label{sec:related}

A number of efforts have been made to characterize and
  understand failures of individual hardware components (e.g., 
  disks~\cite{Schroeder:2007,Schroeder:2010,Pinheiro:2007,Bairavasundaram:2007,Bairavasundaram:2008,Bairavasundaram:2008:2},
  memory~\cite{Narayanan:2016,Meza:2015,Schroeder:2016,Schroeder:2009,Sridharan:2015},
  and others~\cite{Jiang:2008})
  as well as file systems~\cite{Ganesan:2017,Rubio:2009,Lu:2013,Alagappan:2016,Prabhakaran:2005,Gunawi:2008,Pillai:2014}).
Compared with failure studies of individual storage system components,
little prior work analyzed reliability of distributed file and storage systems.
Ford et al.~\cite{Ford:2010} characterizes the availability of Google's storage systems,
  with a focus on correlated failures.
Our study is fundamentally different and complementary to the prior work. 
Our study characterizes the manifestation of both component failures 
    and resource overload as performance issues.

\sysname{} is built upon the wealth body of work on failure detection~\cite{Leners:2011,
    Leners:2015,Huang:2018,Hayashibara:2004,Chandra:1996,Aguilera:2002,Huang:2017}.
In \S\ref{sec:comparison}, we discuss the failure patterns that cannot be handled 
    by the state-of-the-art failure detection methods.
Panorama~\cite{Huang:2018} enhances observability to detect gray failures
  by inserting failure
  reporting code at {\it observation points} in the software programs.
\sysname{} shares the same insight as
  Panorama that the ability of observing the system from the
  viewpoint of the clients is a necessity for monitoring complex distributed
  systems. On the other hand, we show that certain failure patterns 
  cannot be detected by Panorama-like approach if they do not 
  trigger exceptions in the client code.
\sysname{} proactively probes the system to build observability with the goal 
    of preventing client-perceived issues in the first place.

Active measurement has been used for networks monitoring and
  fault localization~\cite{Zhang:2018,Guo:2015,Tan:2019}.
Pingmesh~\cite{Guo:2015} asks every server in a data center
  to ping each other and
  uses the aggregated ping data for network latency analysis.
NetBouncer~\cite{Tan:2019} leverages the IP-in-IP protocol supported
  by modern switches to actively probe selected network paths
  to pinpoint the faulty links and devices.
  Deepview~\cite{Zhang:2018} leverages monitoring data to diagnose
    virtual disk failures by triaging the root causes into compute,
    network, or storage tiers (it treats the entire storage cluster as a black box).
\sysname{} fundamentally differs from the aforementioned methods in at least two aspects:   
   1) \sysname{} is the first effort for monitoring large-scale distributed file
  systems and the underlying storage infrastructure;
   2) \sysname{} is able to differentiate reliability issues and performance issues---as 
    discussed in \S\ref{sec:comparison}, none of the network monitoring approach considers resource overload and contention;
and 3) \sysname{} goes beyond a failure localization tool but can further pinpoint the root causes
    inside the unhealthy components.

\section{Conclusion}
\label{sec:conclusion}

Our study shows that reliability and performance are inter-related as 
    component failures and resource contention both 
    lead to I/O timeouts or slowdown which is hard to disambiguate. 
This paper advocates the need for identifying and diagnosing resource overload and reliability failures jointly 
    to effectively coordinate recovery strategy. 
We build \sysname{} and deploy it on a peta-scale production system.
Our evaluation and experience show that \sysname{} is effective in 
    providing live forensic support for large-scale distributed systems 
    with negligible overhead.

\section{Acknowledgement*}
We thank  Larry Kaplan (Cray) and Gregory Bauer (NCSA) for having many insightful conversations.

This material is based upon work supported by the U.S. Department of Energy, Office of Science, Office of Advanced Scientific Computing Research, under Award Number 2015-02674. This work is partially supported by NSF CNS 13-14891, and an IBM faculty award. 

This research is part of the Blue Waters sustained-petascale computing project, which is supported by the National Science Foundation (awards OCI-0725070 and ACI-1238993) and the state of Illinois. Blue Waters is a joint effort of the University of Illinois at Urbana-Champaign and its National Center for Supercomputing Application. 

{\normalsize
	\bibliographystyle{acm}
	\bibliography{bibliography.bib}
}

\end{document}